\documentclass{article}

\usepackage{PRIMEarxiv}

\usepackage[utf8]{inputenc} 
\usepackage[T1]{fontenc}    
\usepackage{hyperref}       
\usepackage{url}            
\usepackage{booktabs}       
\usepackage{amsfonts}       
\usepackage{nicefrac}       
\usepackage{microtype}      
\usepackage{lipsum}
\usepackage{fancyhdr}       
\usepackage{graphicx}       
\graphicspath{{media/}}     
\usepackage{amsmath} 
\usepackage{float}
\usepackage{caption}
\usepackage{subcaption}
\usepackage{xcolor}
\usepackage{multirow}
\usepackage{url}
\title{A Comprehensive Analysis of Machine Learning Based File Trap Selection Methods to Detect Crypto Ransomware
}

\author{
  P. Mohan Anand, Hrushikesh Chunduri, Sandeep K Shukla\\
  Department of Computer Science and Engineering \\
  Indian Institute of Technology \\
  Kanpur, INDIA \\
  \texttt{\{pmohan, hrushicnv, sandeeps\}@cse.iitk.ac.in} \\
   \And
   P.V. Sai Charan \\
   New York University \\
   USA \\
  \texttt{v.putrevu@nyu.edu} \\
}

\begin{document}
\maketitle

\begin{abstract}
The use of multi-threading and file prioritization methods has accelerated the speed at which ransomware encrypts files. To minimize file loss during the ransomware attack, detecting file modifications at the earliest execution stage is considered very important. To achieve this, selecting files as traps and monitoring changes to them is a practical way to deal with modern ransomware variants. This approach minimizes overhead on the endpoint, facilitating early identification of ransomware. This paper evaluates various machine learning-based trap selection methods for reducing file loss, detection delay, and endpoint overhead. We specifically examine non-parametric clustering methods such as Affinity Propagation, Gaussian Mixture Models, Mean Shift, and Optics to assess their effectiveness in trap selection for ransomware detection. These methods select M files from a directory with N files (M<N) and use them as traps. In order to address the shortcomings of existing machine learning-based trap selection methods, we propose APFO (Affinity Propagation with File Order). This method is an improvement upon existing non-parametric clustering-based trap selection methods, and it helps to reduce the amount of file loss and detection delay encountered. APFO demonstrates a minimal file loss percentage of 0.32\% and a detection delay of 1.03 seconds across 18 contemporary ransomware variants, including rapid encryption variants of lock-bit, AvosLocker, and Babuk.

\end{abstract}


\section{Introduction}
\begin{figure}[h!]
 \centering
 \includegraphics[width=0.7\textwidth]{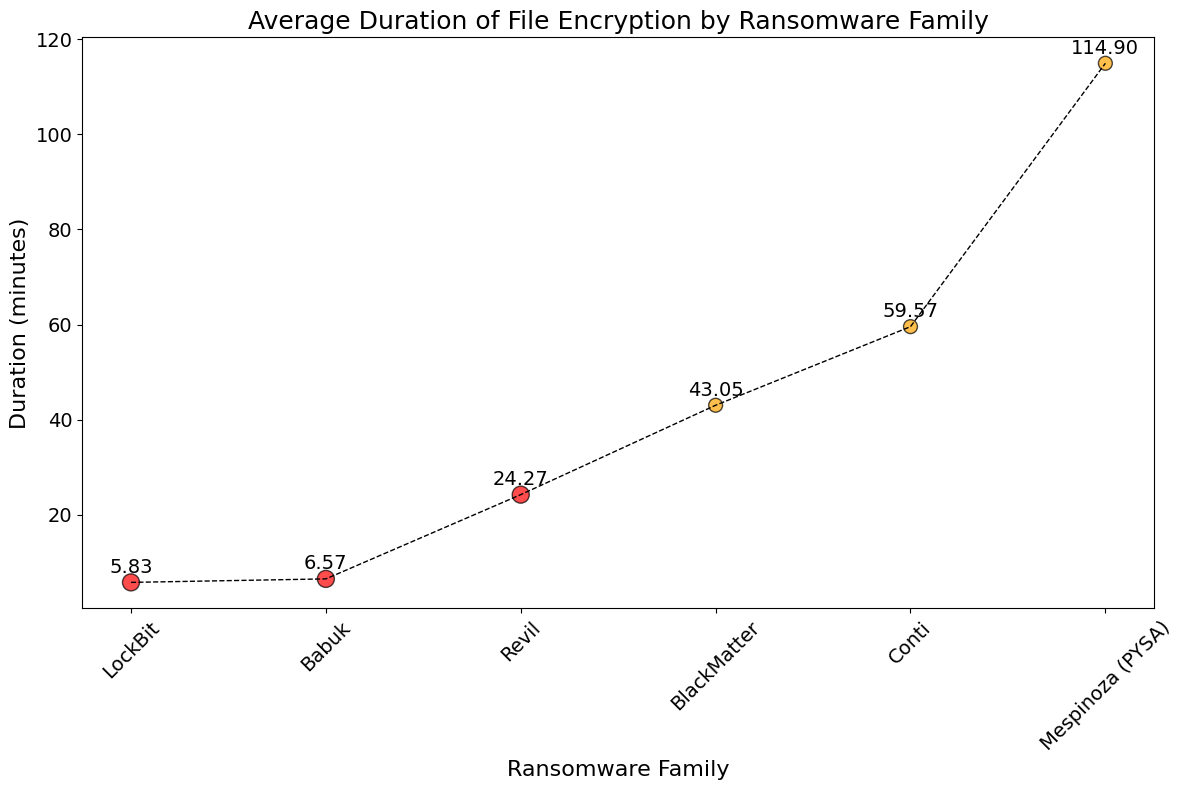}
 \caption{Average duration of file encryption by multiple ransomware variants on 98,561 files totaling 53 GB - Reported by Splunk \cite{speedtest}}
 \label{AES}
\end{figure}

Crypto ransomware is a pernicious malware that aims to encrypt files on any endpoint and subsequently demand a substantial ransom in exchange for the decryption key. Recent advancements in multithreading and intermittent encryption have allowed the encryption of files in parallel and only specific sections of files to be encrypted, resulting in a significant increase in file loss during ransomware attacks \cite{cyberarc}. For example, LockBit 3.0 variants can encrypt 200,000 files in 7 minutes, whereas the Rorschach variant achieves the same in 4.5 minutes \cite{rorschach}. In addition, researchers from Splunk assessed the encryption speed of various ransomware variants on 98,561 files totaling 53 GB, with their findings highlighting the rapid encryption rates of the LockBit and Babuk variants, as shown in Figure \ref{AES} \cite{speedtest}. The swift encryption rates showcased by these variants underscore the critical role of early detection in mitigating file loss and latency during a ransomware attack.

Detecting ransomware is difficult for static malware analysis methods because modern variants use packers and obfuscation techniques, complicating the detection process \cite{moser2007limits,poudyal2019pefile,charan2022dotmug}. Static methods analyze strings, opcodes, and file section metadata extracted from the binary of ransomware to build models for identifying signatures. However, with recent trends, ransomware uses polymorphism techniques to mutate its structure into new versions and assist in crafting many different ransomware variants with the existing code \cite{charan2023text}. Mutation effects, such as code alteration and structure modification, have been observed in numerous ransomware variants. For instance, cybercriminals leverage leaked code from Babuk variants to create new ransomware variants like Play, Cylance, and Rorschach \cite{mutation}. This customization makes it more difficult for static and signature-based methods to detect these ever evolving ransomware attacks.

The majority of dynamic analysis methods rely on API call sequences obtained during ransomware execution to build ML models to aid ransomware detection \cite{sgandurra2016automated,chen2017automatic,vinayakumar2017evaluating,kok2019prevention,kok2020evaluation,hampton2018ransomware,anand2022comprehensive}. However, in real-time scenarios, obtaining API call sequences can introduce a delay in detection, which may lead to substantial file loss. In addition, modern ransomware variants, such as Magniber, explore direct system call methods to bypass API call sequences \cite{DirectSysCall}. This makes it even more challenging for API call-based dynamic detection methods to effectively detect and terminate ransomware attacks. The use of hardware performance counters to detect ransomware in its earliest phases of execution has drawn attention because ransomware has shown no interaction with hardware registers to manipulate or evade detection \cite{kadiyala2020hardware,bahador2019hlmd,olani2022deepware,alam2020rapper,anand2023hiper,putrevu2023early}. Yet, using HPC registers for ransomware detection on endpoints is not ideal, as it is prone to false positives when multiple programs run concurrently. 

In the early ransomware detection domain, deception using trap files is an emerging area of research. This approach selects trap files on a specific endpoint and tracks changes in these traps to identify the ransomware's malicious activity. This approach's success depends on how early the ransomware targets selected trap files at a given endpoint. The selection of trap files plays a vital role in the early detection of ransomware. When ransomware targets a trap file after encrypting numerous other files on endpoints, it leads to significant file loss. In contrast, if the ransomware hits the selected trap at the earliest, the file loss will be minimal, proving to be very effective even with faster ransomware variants. Modern ransomware variants are capable of launching parallel threads to encrypt files across multiple folders, as well as performing encryption in a random order. Therefore, relying solely on file traps set up in a few selected directories and using a heuristic approach may result in increased file loss. This behavior of modern ransomware necessitates the use of a non-heuristic approach for trap selection.

In this paper, we discuss the application of machine learning (ML) to selecting files as traps using a non-heuristic, data-driven approach. These ML models select the potential files (M) from each directory containing N files as traps (where M<N) and monitor for any modifications to these M files to alert and terminate the ransomware from causing further file loss. This is done by implementing non-parametric clustering methods on the file parameters in every directory where cluster centers are identified as representative files for a group of clusters. Non-parametric clustering is a method of grouping data points without assuming a specific distribution or number of clusters beforehand. Selecting non-parametric clustering methods for trap identification is essential to avoid using rule-based methods when deciding the number of groups in a directory. In any given directory, the number of clusters to be formed should be entirely based on parameters such as file size, file type, file creation, and update date, but not on user input. Since the number and characteristics of files differ across directories, setting the cluster number based on user input is not effective for trap selection. Therefore, we chose non-parametric clustering algorithms to identify potential trap files in each directory. We consider the below four widely used non-parametric clustering algorithms as our candidate ML-based clustering models
\begin{enumerate}
    \item Affinity Propagation
    \item Gaussian Mixture Models
    \item Mean Shift
    \item Optics (Ordering points to identify the clustering structure)
\end{enumerate}

and evaluated the effectiveness of these models based on the parameters of file loss, detection delay, and memory consumption across 18 ransomware variants that carry out encryption faster. Furthermore, we have highlighted the reasons and scenarios where the above-mentioned models have resulted in higher file loss and latency in detection during our evaluation. In order to reduce the file loss and detection delay, we have implemented APFO (Affinity Propagation with File Order), which enhances the current machine learning-based clustering methods to overcome the shortcomings of the existing approaches. APFO exhibits a minimal 0.32\% rate of file loss and a detection delay of 1.03 seconds when tested against 18 ransomware variants, including fast encryption variants such as lock-bit, AvosLocker, and Babuk.

\subsection{The major contribution of this work includes:}
\begin{enumerate}
    \item We evaluate the machine learning based trap file selection algorithms and highlight the advantages and drawbacks of each method by considering real-time endpoint deployment scenarios. This research objective answers to the following questions:
    \begin{itemize}
        \item We evaluate file loss, detection delay, and memory overhead by running eighteen distinct ransomware families in a controlled environment, where trap files are selected using machine learning-based methods.
        \item How well do these models handle faster-encrypting ransomware variants?
    \end{itemize}
    \item We propose APFO (Affinity Propagation with File Order) to minimize file loss and detection delays across all ransomware variants. We discuss the following points under this research objective:
    \begin{itemize}
        \item Under what scenarios do the existing ML-based file trap selection methods face delayed detection?
        \item Given the dynamic nature of ransomware, can future variants evade APFO, resulting in delayed detection and increased file loss?
    \end{itemize}     
\end{enumerate}

The rest of the paper is organized as follows: Section 2 reviews related work on the use of file traps for early ransomware detection, highlighting the advantages and limitations of each approach. Section 3 provides background information on non-parametric clustering methods relevant for the trap selection. Section 4 outlines the methodology for selecting traps using non-parametric clustering methods at any given endpoint. In Section 5, we compare and discuss various ML-based trap selection methods to identify the most suitable approach for minimizing file loss and detection delay during ransomware execution. Finally, Section 6 summarizes the key findings and contributions of this research, as well as outlines future work in this area.

\section{Related Works}
Using trap files as a deceptive technique to identify ransomware has gained significant interest in recent years. The authors of related research explored various ways, including heuristic and non-heuristic approaches, to select or deploy trap files for early detection. Heuristic approaches involve deriving solutions from practical experience or knowledge. For example, if one is aware that a particular ransomware variant utilizes depth-first search for directory traversal during encryption, heuristic methods can strategically place traps following the same order. Conversely, non-heuristic methods rely on system procedures and algorithmic implementations to reach a solution. 
\newline


\begin{table}[]
\centering
\scalebox{0.85}{
\resizebox{\textwidth}{!}{%
\begin{tabular}{@{}|p{1.8cm}|p{8cm}|p{7cm}|@{}}
\toprule
\small \textbf{Method Name} &
  \small \hspace{2cm}\textbf{Trap File Selection Criteria} &
  \small \hspace{3cm}\textbf{Takeaways} \\ \midrule
\small Lee et al. \cite{lee2017make} &
  \small \begin{itemize}
    \item Create files in user-specific folders like Desktop and Documents with names starting in alphabetical and reverse alphabetical order to attract ransomware.
    \item File size parameter consideration: Multiple small trap files and one large file are generated in the selected directories.
  \end{itemize} &
  \small \begin{itemize}
    \item Delayed detection in cases of random order encryption.
    \item Do not select all folders to place traps.
    \item Delayed detection with ransomware variants that use multithreading to encrypt files.
  \end{itemize} \\ \midrule
\small RLocker \cite{gomez2022inhibiting} &
  \small \begin{itemize}
    \item Creates symbolic links to a central FIFO file in every directory.
    \item Symbolic link files are named in alphabetical order.
    \item Symbolic link files are marked as hidden.
  \end{itemize} &
  \small \begin{itemize}
    \item Delayed detection in cases of random order encryption.
    \item Cerber ransomware evades this detection method by considering file size parameter before it starts encryption.
  \end{itemize} \\ \midrule
\small R-Sentry \cite{sheen2022r} &
  \small \begin{itemize}
    \item Ranks the directories based on file access patterns, size, and type.
    \item Traps are placed in the top-ranked directories and the root user directory.
  \end{itemize} &
  \small \begin{itemize}
    \item Delayed detection in cases of random order encryption.
    \item Do not select all folders to place traps.
    \item Delayed detection with ransomware variants that use multithreading to encrypt files.
  \end{itemize} \\ \midrule
\small Yun Feng et al. \cite{feng2017poster} &
  \small \begin{itemize}
    \item Uses predetermined paths targeted by ransomware families to place traps.
    \item Creates specific types (e.g., txt, doc, pdf) of traps to attract ransomware.
    \item Uses the API hooking approach for file access based on Windows APIs.
  \end{itemize} &
  \small \begin{itemize}
    \item Do not select all folders to place traps.
    \item Delayed detection with ransomware variants that use multithreading to encrypt files.
    \item Ransomware evades this approach through direct system calls, bypassing Windows APIs.
  \end{itemize} \\ \midrule
\small RWGuard \cite{mehnaz2018rwguard} &
  \small \begin{itemize}
    \item Trap files are created in every directory.
    \item Users can specify the number of traps to be placed in every directory.
    \item File types and file sizes are considered while generating new traps.
  \end{itemize} &
  \small \begin{itemize}
    \item Increased system resource consumption due to the creation of additional files.
    \item Ransomware can skip the newly generated files based on the created date property.
  \end{itemize} \\ \midrule
\small RTR-Shield \cite{anand2023rtr} &
  \small \begin{itemize}
    \item Selecting traps is done based on understanding the file access patterns used by various ransomware families.
    \item Considered Depth First and Breadth First order of file traversal.
    \item Consider the alphabetical and reverse alphabetical ordering of file names for trap selection.
  \end{itemize} &
  \small \begin{itemize}
    \item Do not select all folders to place traps.
    \item Delayed detection with ransomware variants that use multithreading to encrypt files.
    \item Delayed detection in cases of random order encryption.
  \end{itemize} \\ \midrule
\small CryptoStopper \cite{CryptoStopper} &
  \small \begin{itemize}
    \item Trap files are generated with random names, extensions, and sizes.
    \item Generated trap files are marked as hidden.
    \item Selection criteria details are not explicitly mentioned as it is a proprietary paid tool.
  \end{itemize} &
  \small \begin{itemize}
    \item Increased system resource consumption due to the creation of additional files.
    \item Ransomware can skip the newly generated files based on the created date property.
  \end{itemize} \\ \midrule
\small RTrap \cite{ganfure2023rtrap} &
  \small \begin{itemize}
    \item Trap files are selected from every directory.
    \item Employed Affinity Propagation (a non-parametric clustering algorithm) to select traps across directories.
    \item Consider file size, creation date, update date, and file type as features for performing clustering.
  \end{itemize} &
  \small \begin{itemize}
    \item Non-heuristic approach.
    \item The approach needs to be tested on alphabetic and reverse alphabetical ransomware variants.
    \item The approach's effectiveness on endpoints with larger directories (more than 200 files in a single folder) has to be evaluated.
  \end{itemize} \\ \bottomrule
\end{tabular}%
}
}
\caption{Summary of related works}
\label{SUMMARYRW}
\end{table}

Lee et al. \cite{lee2017make} proposed a strategy to create files in user-specific folders like desktop and documents with names starting in alphabetical order to attract ransomware. Multiple small trap files and one large file are generated in the selected directories to counteract encryption based on file size and access order. In another approach, Sinha et al. \cite{sheen2022r} proposed R-Sentry, which  involves ranking folders based on various criteria that ransomware might use when searching for files to encrypt. These criteria include factors such as the most recently accessed files, file size, creation and last accessed times, frequency of access, and common file types within folders. Trap files are then placed in the top-ranked folders, specifically in root directories, to maximize the chances of detection. Similarly, Mohan et al. \cite{anand2023rtr} proposed RTR Shield, where trap selection is done based on understanding the file access patterns used by various ransomware families, which often employ depth-first and breadth-first traversal orders for encryption. In addition, they considered the alphabetical ordering of files in the selected directories to be a trap. These approaches \cite{lee2017make,sheen2022r,anand2023rtr} are ineffective for early detection when ransomware variants encrypt files in random order. Moreover, these methods face challenges with ransomware variants that use multithreading to encrypt files in multiple folders simultaneously.
\newline

In a different work, Jose et al. \cite{gomez2018r,gomez2022inhibiting} proposed R-Locker, which involves creating symbolic links to a central FIFO file in each folder of the filesystem to maximize protection. FIFOs exhibit a unique property where if a process attempts to read from an empty FIFO, it will be blocked until a writing process is available. This characteristic is leveraged to stop ransomware in its tracks from accessing a trapfile, effectively preventing it from proceeding with file encryption. Moreover, the symbolic links are strategically named to be alphabetically first and can be marked as hidden to remain unnoticeable to users, increasing the likelihood of being accessed by ransomware during its file scanning process. However, Cerber ransomware evades this method by targeting only files meeting a minimum size threshold for encryption, leading it to evade the trap files created by R-Locker, which have a size of 0 bytes. As a result, Cerber can execute without triggering the detection mechanism of R-Locker.
\newline

Yun Feng et al. \cite{feng2017poster} proposed a trap file creation approach that involves creating specific types of trap files (e.g., txt, doc, pdf) in predetermined paths that are commonly targeted by ransomware. The proposed method hooks the Windows APIs \textit{FindFirstFile} and \textit{FindNextFile} to ensure ransomware targets these trap files first, monitoring for malicious behavior like encryption by analyzing changes in file randomness or file type modification. Upon detecting such behavior, the malicious activity is stopped to prevent further damage. However, modern ransomware avoids these detection methods by directly using system calls, thereby bypassing the API calls that the proposed method relies on. In another work, Mehnaz et al. \cite{mehnaz2018rwguard} proposed RWGuard, which involves an automated trap generator as one of its behavior monitoring components. The trap generator creates trap files in a directory that mimic the names of the original files by appending user-defined prefixes or suffixes to the filenames based on user settings. Users can customize the number of trap files per directory. The traps are generated with various extensions and content derived from neighboring files to ensure they appear legitimate. 
\newline

Similarly, Watchpoint released a tool called \textit{Cryptostopper} to detect ransomware using the trap file approach \cite{CryptoStopper}. The trap file selection strategy in CryptoStopper involves creating "Watcher Files" that are designed to mimic legitimate files within protected directories. These watcher files are generated with random names, extensions, and sizes, allowing them to blend seamlessly with actual data. This randomness helps to confuse ransomware, making it more likely to interact with the trap files.  Additionally, the hidden attribute of these files prevents users from seeing them, reducing the risk of accidental deletion and ensuring continuous monitoring for ransomware activity. However, the technical details on the methods of creating the trap files are not disclosed as it is a proprietary paid tool. The above discussed approaches \cite{feng2017poster,mehnaz2018rwguard,CryptoStopper} generate new trap files at the end points to lure the ransomware. But this enables ransomware to skip the newly generated files based on the created date property, and it may also lead to the consumption of additional system resources, as new files are generated through this method. 
\newline

All the methods discussed above depend on heuristics in order to select the traps. The increasing complexity of newer ransomware variants, which employ multi-threaded and random order file encryption poses challenges for timely detection using existing trap selection methods. In response to the limitations of heuristic approaches, Gadisa et al. \cite{ganfure2023rtrap} proposed RTrap as a non-heuristic method that utilizes machine learning techniques for trap selection.  RTrap's trap file selection strategy utilizes a data-driven approach with the Affinity Propagation (AP) algorithm, enabling clustering of user files without the need for an initial estimation of the required trap files. Through iterative exchange of attractiveness messages among files, the model in RTrap identifies a subset of files that serve as trap files representing the entire file collection on a given end point. Subsequently, a decoy-watcher actively monitors the generated traps in real-time to detect any unauthorized access, facilitating prompt defensive actions against potential ransomware threats. RTrap has demonstrated superior performance compared to traditional heuristic-based methods by significantly reducing file loss and minimizing detection delays. Table \ref{SUMMARYRW} includes a summary of related works, briefing on trap selection criteria, and takeaways.
\newline

The authors of RTrap used affinity propagation, a non-parametric clustering algorithm as part of their methodology to select traps for ransomware detection. Nevertheless, it is essential to validate other non-parametric clustering algorithms to enhance trap selection techniques for early detection of ransomware threats. Our objective was to assess the effectiveness of popular non-parametric clustering algorithms, such as Affinity Propagation, Gaussian mixture models, Mean Shift, and Optics, in trap selection for early detection of ransomware. 

\section{Background}
Clustering is an approach followed to group similar data points based on the attributes or features of the data points. Some of the widely known clustering algorithms, such as K-means or K-medoids, require explicit mention of the parameter ‘K’ before grouping the data points. In those methods, we explicitly mention the number of groups we want before the clustering begins. Based on the mentioned K value, the clustering algorithms will group the data points into respective clusters. After forming the clusters, we can identify the cluster centers, which are considered to be representative data points for the specific clusters. 

 Based on our problem statement, we need to select a few representative files from every directory to be chosen as traps to effectively detect and combat ransomware attacks. We follow this principle of selecting traps from every directory to be resilient and quick against the ransomware variant that launches parallel threads to encrypt more than one directory at a given point in time, ensuring a proactive defense strategy. 

In the analogy with clustering algorithms, the representative files can be likened to the cluster centers, which are considered the focal points of each cluster. However, we cannot fix the number of clusters to be formed beforehand because each directory is different in terms of its structure. The structure of a directory refers to the number of files present, types of files, creation date, update date, and also the size of the files. 

The requirement is to use the clustering algorithms which can automatically select the number of clusters based on the properties of the files present inside each directory rather than on the user input. For this reason, we chose a \textbf{non-parametric} clustering approach that does not require specifying the ‘K’ value before grouping the data points, allowing for a more flexible and adaptive clustering process. These non-parametric clustering methods are designed to automatically select the best possible ‘K’ groups.

 In our work, we evaluated four widely used non-parametric clustering algorithms 1. Affinity Propagation 2. Mean Shift 3. Gaussian Mixture Models and 4. Optics to determine the most effective method for selecting traps in ransomware detection. We discuss the functionality of each of these methods in the subsections below to provide insights into how each algorithm contributes to the selection of traps for ransomware detection.

\subsection{Affinity Propagation}
Affinity Propagation is a clustering algorithm that finds exemplars in data points and creates clusters based on them \cite{dueck2007non}. Exemplers are input set data points, which represent clusters. Unlike traditional clustering algorithms, including k-means, which require the number of clusters to be specified beforehand, Affinity Propagation infers the number of clusters from the data. The algorithm operates by exchanging messages between data points until a suitable set of exemplars and clusters emerges. Affinity propagation exchanges messages of two types: responsibility and availability.

The responsibility $r(i,k)$ indicates how well-suited point $k$ is to serve as the exemplar for point $i$, calculated as:

$$r(i,k) = s(i, k) - \max_{k' \neq k} \{a(i, k') + s(i, k')\}$$

where $s(i,k)$ is the similarity between point $i$ and point $k$. It is the negative squared Euclidean distance and calculated as follows:
\[ s(i, k) = -\|x_i - x_k\|^2 \]

The availability $a(i,k)$ reflects how appropriate it would be for point $i$ to choose point $k$ as its exemplar, given the other points’ preferences for $k$. It is calculated as:

$$a(i, k) = \min \left\{ 0, r(k, k) + \sum_{i' \not\in \{i, k\}} \max \{0, r(i', k)\} \right\}$$

For $k=i$, the self-availability is:

$$a(k, k) = \sum_{i' \neq k} \max \{0, r(i', k)\}$$

Initially, the responsibility and availability matrices are set to zero. It uses input in the form of a similarity matrix, usually coming from negative squared Euclidean distances. Through iterative updates of responsibilities and availabilities, the algorithm converges to a solution in which data points naturally cluster around exemplars. At the end, the algorithm selects exemplars based on the final responsibility and availability values, grouping each point with the exemplar that maximizes the sum of these values. Affinity Propagation has been used in different applications with complex datasets to discover intrinsic structures without parameter tuning.

\subsection{Mean Shift Clustering}
Mean Shift Clustering is a non-parametric iterative clustering technique for identifying dense areas in the feature space; it is also called modes. It works by shifting data points iteratively towards the mean of data points in their neighborhood, hence "shifting" them towards the area of greatest density \cite{comaniciu2002mean}. Essentially, the method considers data points as a representation of the data distribution and aims to locate the highest concentration areas within it. The algorithm starts by defining a kernel function to be used in the calculation of the density estimate; normally, a Gaussian kernel is used. The mean shift vector for each data point $x_i$ is now calculated using the following formula:
\newline
$$ m(x_i) = \frac{\sum_{j=1}^{n} K(x_j - x_i) \cdot x_j}{\sum_{j=1}^{n} K(x_j - x_i)} - x_i $$
\newline
The formula calculates a weighted average of the positions of the neighboring points for $x_i$ , giving a direction and distance (the mean shift vector) to move towards. Here, $K$ stands for the kernel function, $x_j$ for the data points near $x_i$, and $n$ for the total number of data points. The kernel function K($x_j$ – $x_i$) takes into account how far away each point $x_j$ is from $x_i$, with closer points having higher weights. Next, shift the data point \( x_i \) in the direction of the mean shift vector \( m(x_i) \) using:
\[ x_i \leftarrow x_i + m(x_i) \] The above step updates $x_i$ to a new position that is closer to the region with a higher density of data points. This process is repeated until convergence, which is achieved when the points cease to move significantly. After all points have converged to their modes, the points that have converged to the same mode are grouped together to form clusters. This procedure results in the identification of clusters that represent the concentrated regions in the feature space. Mean shift clustering is advantageous as it automatically determines the number of clusters without prior input and can handle clusters of different shapes, making it versatile for diverse applications.

\subsection{Gaussian Mixture Models (GMM) }
In contrast to hard clustering methods, Gaussian Mixture Models (GMM) allow a probabilistic model in which every data point belongs to multiple clusters with different probabilities. In GMM, it is assumed that the data is generated from a mixture of several Gaussian distributions, each representing a cluster. The probability density function of GMM is as follows:

\[ p(x) = \sum_{k=1}^{K} \pi_k \mathcal{N}(x | \mu_k, \Sigma_k) \]

where \( \pi_k \) is the mixing coefficient for the \( k \)-th Gaussian component, \( \mu_k \) is the mean vector, \( \Sigma_k \) is the covariance matrix, and \( \mathcal{N}(x | \mu_k, \Sigma_k) \) is the multivariate Gaussian distribution, which is defined as:

\[ \mathcal{N}(x | \mu_k, \Sigma_k) = \frac{1}{(2\pi)^{d/2} |\Sigma_k|^{1/2}} \exp\left(-\frac{1}{2} (x - \mu_k)^\top \Sigma_k^{-1} (x - \mu_k)\right) \]

In GMM, the parameters such as \(\pi_k\), \(\mu_k\), and \(\Sigma_k\) are estimated through the Expectation-Maximization (EM) algorithm \cite{dempster1977maximum}. The EM method iteratively optimizes the parameters to maximize the likelihood of the observed data. However, an essential aspect of Gaussian Mixture Models (GMM) is determining the ideal number of clusters represented by \(K\). Model selection criteria, such as the Akaike Information Criterion (AIC) and the Bayesian Information Criterion (BIC), are used to determine the most suitable number of clusters \(K\) \cite{akaike1998information,schwarz1978estimating}.

The AIC is defined as:

\[ \text{AIC} = 2k - 2\ln(\hat{L}) \]

where \(k\) is the total number of parameters, which includes the means, covariances, mixing coefficients, etc., and \(\ln(\hat{L})\) is the log-likelihood of the model.

The BIC is defined as:

\[ \text{BIC} = k\ln(n) - 2\ln(\hat{L}) \]

where \(n\) is the number of data points. Both AIC and BIC criteria penalize models with a higher number of parameters to prevent overfitting. The optimal number of clusters \(K\) is selected by fitting GMMs with various values and selecting the one that yields the lowest AIC or BIC value. In this scenario, the clustering approach becomes non-parametric, as it relies on data-driven criteria rather than predefined assumptions about the number of clusters \(K\). This flexibility makes it possible for GMMs to automatically adapt to the underlying structure of the data, making them useful for our scenario.

\subsection{Optics}
Optics (Ordering Points To Identify Clustering Structure) is a clustering algorithm that can group data of varying densities \cite{ankerst1999optics}. In contrast to usual clustering methods that require a set number of clusters or density thresholds, Optics does non-parametric clustering by creating a reachability plot that shows the hierarchical structure of clusters in the data. This method enables to identify clusters of various shapes and sizes, as well as distinguish noise points. Optics groups data points by `\textit{ReachabilityDistance}' to reveal their structure without assuming a fixed number of clusters. 

Optics relies on two key concepts: `\textit{ReachabilityDistance}' and `\textit{CoreDistance}'. For a given data point \( p \), the reachability distance to a point \( q \) is defined as:

\[ \text{Reachability}(p, q) = \max(\text{CoreDistance}(p), \text{Distance}(p, q)) \]

where \(\text{Distance}(p, q)\) is the Euclidean distance between \( p \) and \( q \), and \(\text{CoreDistance}(p)\) is defined as the minimum distance from \( p \) to its \( \text{minPts} \) value. Here, \( \text{minPts} \) refers to the minimum number of samples in a neighborhood for a point to be considered as a core point. In this work, we select the most appropriate  \( \text{minPts} \) for the cluster formation using the DBCV (Density-Based Clustering Validation) method \cite{moulavi2014density}. DBCV is specifically applicable for density-based clustering algorithms like Optics  measure the clustering quality by evaluating the compactness and separation of clusters by tuning with multiple  \( \text{minPts} \) values. Through the process of ordering points according to their reachability distances, Optics is able to construct a reachability plot that represents the clustering structure. The valleys in the reachability plot correspond to dense clusters, while the peaks in the plot indicate noise or regions that are less dense. This method  provides a comprehensive view of the data's structure, enabling the identification of clusters with varying densities and sizes.

\section{Methodology}

\begin{figure}[h!]
 \centering
 \includegraphics[width=\textwidth]{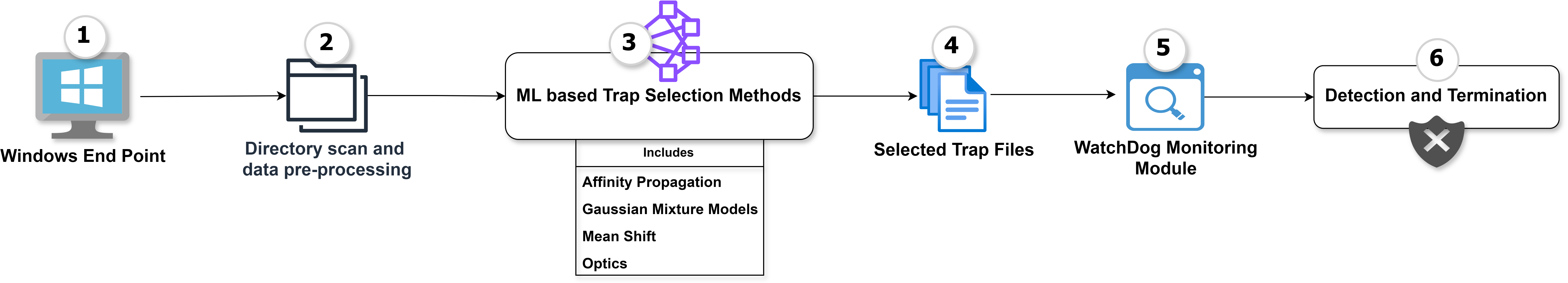}
 \caption{Concept diagram of ML based Trap selection for Ransomware detection}
 \label{CD}
\end{figure}

As shown in Figure \ref{CD}, the process for implementing an ML-based trap file selection to successfully detect and stop ransomware activity is a six-step process. The significance of each step is explained below.
\newline

\textbf{1) End point setup}: \label{sec4ep} In our work, we select Windows endpoints to evaluate methods for choosing traps based on machine learning (ML). This choice is driven by the fact that most ransomware families are designed to target Windows-based systems. The end point is configured with a Windows 10 Pro operating system with 8 GB of RAM and 256 GB of storage. We configured the network of the end point to be in \textit{host only} mode to prevent the network from being affected during the ransomware execution. The setup requires a Python interpreter and several Python packages to execute the trap selection methods based on machine learning (ML) on the endpoint. To ensure everything runs smoothly, we installed all the necessary applications and packages on the endpoint. To complete the end point setup, we have downloaded files from the public repositories of GovDoc \cite{garfinkel2009bringing} and Kaggle \cite{kaggle} and placed the files across multiple directories on the end point. The user data on the end point consists of 46 folders containing 20,558 files, with a total size of 4 gigabytes.
\newline

\textbf{2) Directory Scan and Data Pre-processing }: After setting up the endpoint, we proceed by scanning all directories within the endpoint to identify those with a sufficient number of files. Specifically, we select directories containing three or more files. This step ensures we avoid directories that are either empty or have two or fewer files, as they might not provide enough data for clustering and selecting candidate traps. Additionally, we exclude operating system directories, such as "C:\textbackslash Windows" and "C:\textbackslash Program Files" since ransomware typically does not target these during encryption process. By focusing on all the user directories, we aim to counteract ransomware that uses parallel threading to encrypt multiple directories simultaneously.

Next, we process each directory individually to extract features and pre-process the data before clustering and selecting candidate trap files. For each directory, a dataset is extracted from the files, including features such as file type, file size, file creation time, and file update time. These features are chosen to address ransomware behaviors that target specific file types, sizes, or recently created or modified files.
To ensure the dataset is suitable for clustering, we apply several pre-processing techniques:

\begin{enumerate}
    \item File Size (numerical) : We normalize file sizes using standard scalar normalization \cite{domingos2012few} , which adjusts the data to have a zero mean and unit variance.
    \item File Type (Categorical): We encode file types into integers (e.g., pdf = 0, mp3 = 1).
    \item Creation and Update Times (Date-Time): We apply sine and cosine transformations to these values to capture their cyclic nature of date-time values \cite{datetime}.
\end{enumerate}

After pre-processing, each directory results in an \textit{MxN} dataset, where M represents the number of files in the directory and N represents the number of features obtained after pre-processing. In the following step, we implement the linear dimensionality reduction method PCA on the \textit{MxN} dataset to manage multicollinearity (when features are highly correlated) by converting the correlated features into a set of linearly independent principle components \cite{jolliffe2016principal}. Furthermore, PCA reduces dimensionality, making data points more distinguishable in the reduced feature space, which improves the accuracy of clustering methods such as non-parametric clustering. This standardized dataset is then used in the next step, which involves non-parametric clustering to identify potential trap files.
\newline

\textbf{3) ML based trap selection methods}: We start with an input list of directories and their corresponding \textit{MxN} datasets obtained from the previous step, where M represents the number of files and N represents the number of features. Our approach involves applying non-parametric clustering algorithms to the dataset of a directory to identify potential traps in that directory. This method is continued on all the input lists of directories to select the traps across the end point. We implement four non-parametric clustering algorithms: Affinity Propagation, Gaussian Mixture Models, Mean Shift, and Optics. As discussed in the Background section, these methods do not require the specification of the number of clusters (K) in advance. Each algorithm independently forms clusters within the dataset for each directory. After forming clusters, we identify representative centers for each cluster, which are referred to as traps. Each clustering algorithm generates a list of selected traps specific to that method. For instance, traps identified by affinity propagation can be different from those selected by other methods such as mean shift. Finally, we save the list of traps selected across the end point by each of the four techniques separately.  The most effective trap selection method is evaluated using the trap files selected by these four methods, which are further discussed in section-\ref{sec5}.
\newline

In \textbf{step 4}, the selected trap files are renamed with a user-defined suffix to prevent users from making real-time modifications and maintain the integrity of the trap files. Next, in \textbf{step 5}, the selected trap file list is passed to the file monitoring module to detect any changes like renaming, deletion, or content modification as shown in Figure-\ref{CD}. We used the Python \textit{Watchdog} module to perform the trap file monitoring, as it allows developers to trigger specific actions or processes in response to these changes, enhancing automation and responsiveness in applications \cite{watchdog}. Finally, in \textbf{step 6}, ransomware is detected and terminated when the watchdog raises an alert indicating a potential file modification to any of the selected trap files. Upon the alert, all active processes are stopped to halt ransomware activities and safeguard against file loss. The entire process, from step 2 to step 6, is automated. The system is configured to re-scan the directories within a user-defined time frame, select a new set of traps, and monitor them for any modifications. 

\section{Results and Discussion}
\label{sec5}
In this work, we employed four machine learning-based trap selection methods as outlined in the methodology. Each method generates a list of potential traps for files on a given endpoint. The selected traps are evaluated based on three main criteria: \textbf{File loss, Detection Delay, and Utilization of system resources and performance impact}. 
\begin{table}[htp]
\centering
\resizebox{0.3\columnwidth}{!}{%
\begin{tabular}{|cl|}
\hline
\multicolumn{2}{|c|}{\textbf{Ransomware Families}}                                                                                                                                                                                                                                                   \\ \hline
\multicolumn{1}{|c|}{\textbf{Category-1}}                                                                                                         & \multicolumn{1}{c|}{\textbf{Category-2}}                                                                                                                 \\ \hline
\multicolumn{1}{|l|}{\begin{tabular}[c]{@{}l@{}}Atomsilo\\ AvosLocker\\ Babuk\\ BlackMatter\\ Cerber\\ Lockbit\\ Lorenz\\ Surtr\end{tabular}} & \begin{tabular}[c]{@{}l@{}}Conti\\ Cuba\\ Demonware\\ GlobeImposter\\ Intercobros \\ Karma\\ Magniber\\ Makop\\ Mespinoza\\ Mountlocker\end{tabular} \\ \hline
\end{tabular}%
}
\caption{List of Ransomware Families.
\textbf{Category -1} : Ransomware launches more than 10 threads in its execution. \textbf{Category -2} : Ransomware launches less than 10 threads in its execution.}
\label{RFAM}
\end{table}

For the evaluation, we consider 18 ransomware variants known for their rapid encryption rates and global prevalence. We categorized the ransomware families into two types based on their multi-threaded capabilities. The first category includes ransomware families that initiate more than 10 threads to perform encryption across multiple directories simultaneously. In contrast, the second category comprises ransomware families that operate with fewer than 10 threads. The list of ransomware families considered for evaluation is listed in Table \ref{RFAM}. We execute the ransomware samples against the trap monitor to identify the modifications made by the ransomware on the selected trap files. If a ransomware variant triggers any of the selected traps, we stop the ransomware process and record the file loss and detection delay, respectively. This process is repeated for each ransomware family using each trap selection method individually throughout the evaluation. Below, we present and compare the outcomes of the evaluation for the four machine learning-based trap selection methods.

\subsection{File Loss}
\begin{figure}[htp]
 \centering
 \includegraphics[width=0.80\textwidth,height=75mm]{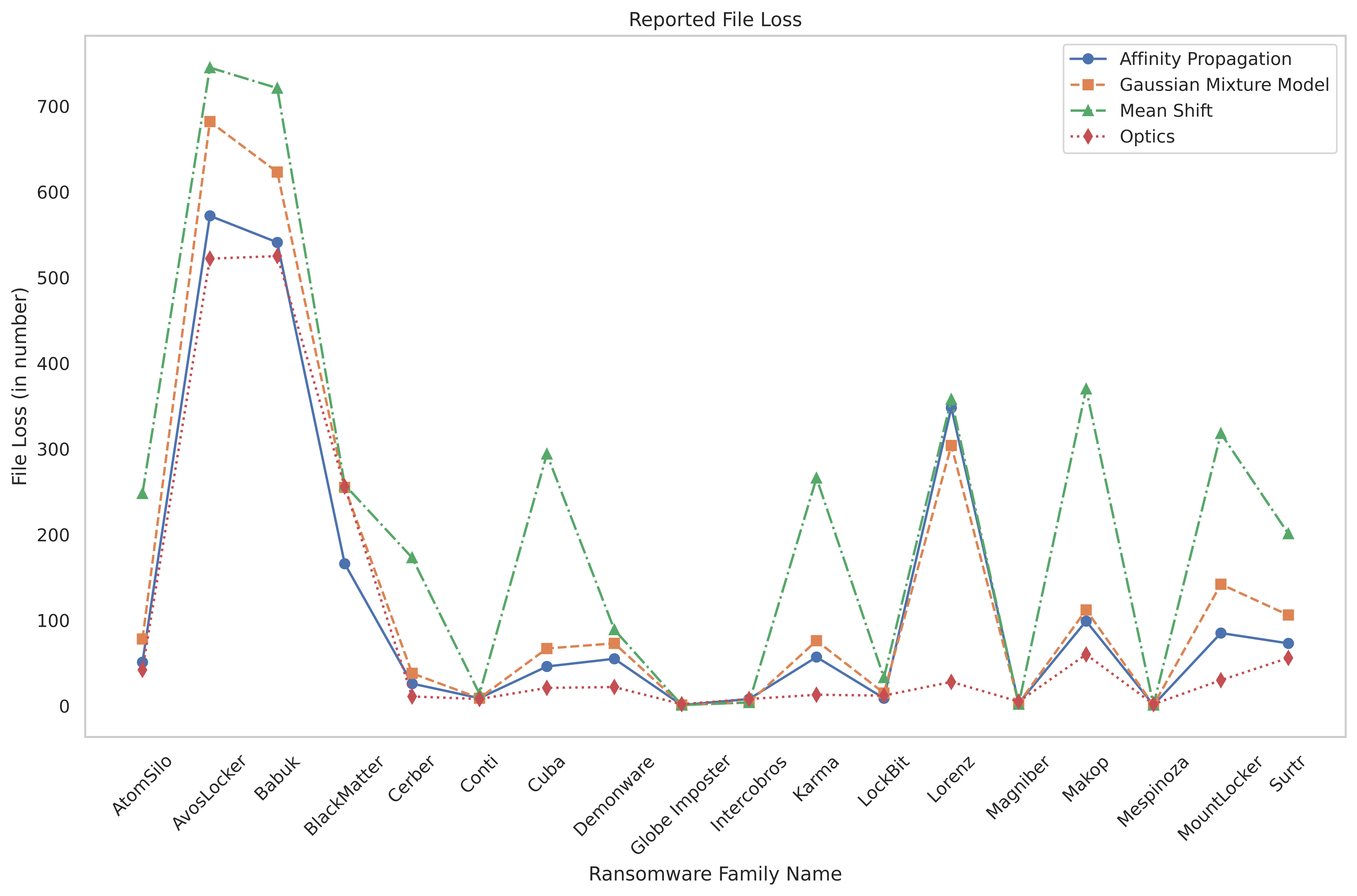}
 \caption{Comparison of ML based trap selection methods (Criteria: File Loss across ransomware variants)}
 \label{FL}
\end{figure}
File loss is measured as the number of files encrypted by the ransomware during its execution. As shown in Figure \ref{CD}, when the ransomware modifies any trap file, an alert is triggered, and the ransomware processes are terminated to prevent further activity. During the encryption, ransomware encrypts the files and adds their own extension to the encrypted file, indicating the encrypted file. For example, Lockbit encrypted files have the extension \textit{.lockbit}, and similarly, AvosLocker encrypts the file with the extension \textit{.avos2}, etc. We define the number of files with these extensions after the ransomware termination as the file loss. We depict the file loss reported for all ransomware variants by employing the four machine learning-based trap selection methods in Figure \ref{FL}. 

During the analysis, we discovered that the ransomware variants AvosLocker and Babuk caused the most file loss across all ML-based trap selection methods, encrypting more than 500 files on the end point containing 20558 files. These ransomware variants use multithreading, faster encryption algorithms, and optimized code to encrypt very large chunks of files in a short span of time. In comparison, we observed that Category-1 ransomware families (mentioned in Table-\ref{RFAM}) result in higher file loss compared to Category-2 due to their multi-threaded nature. In terms of the average file loss across all ransomware families, the Optics method of trap selection showed the best outcome, with 90 files lost during ransomware execution. In contrast, the Mean Shift method resulted in a loss of 250 files, indicating its inefficiency in dealing with faster ransomware variants. In comparison, the Affinity Propagation and GMM trap selection methods resulted in an average file loss of 133 and 155 files, respectively.

\subsection{Detection Delay}
\begin{figure}[htp]
    \centering
    \begin{subfigure}[t]{\textwidth}
        \centering
        \includegraphics[width=0.80\textwidth, height=70mm]{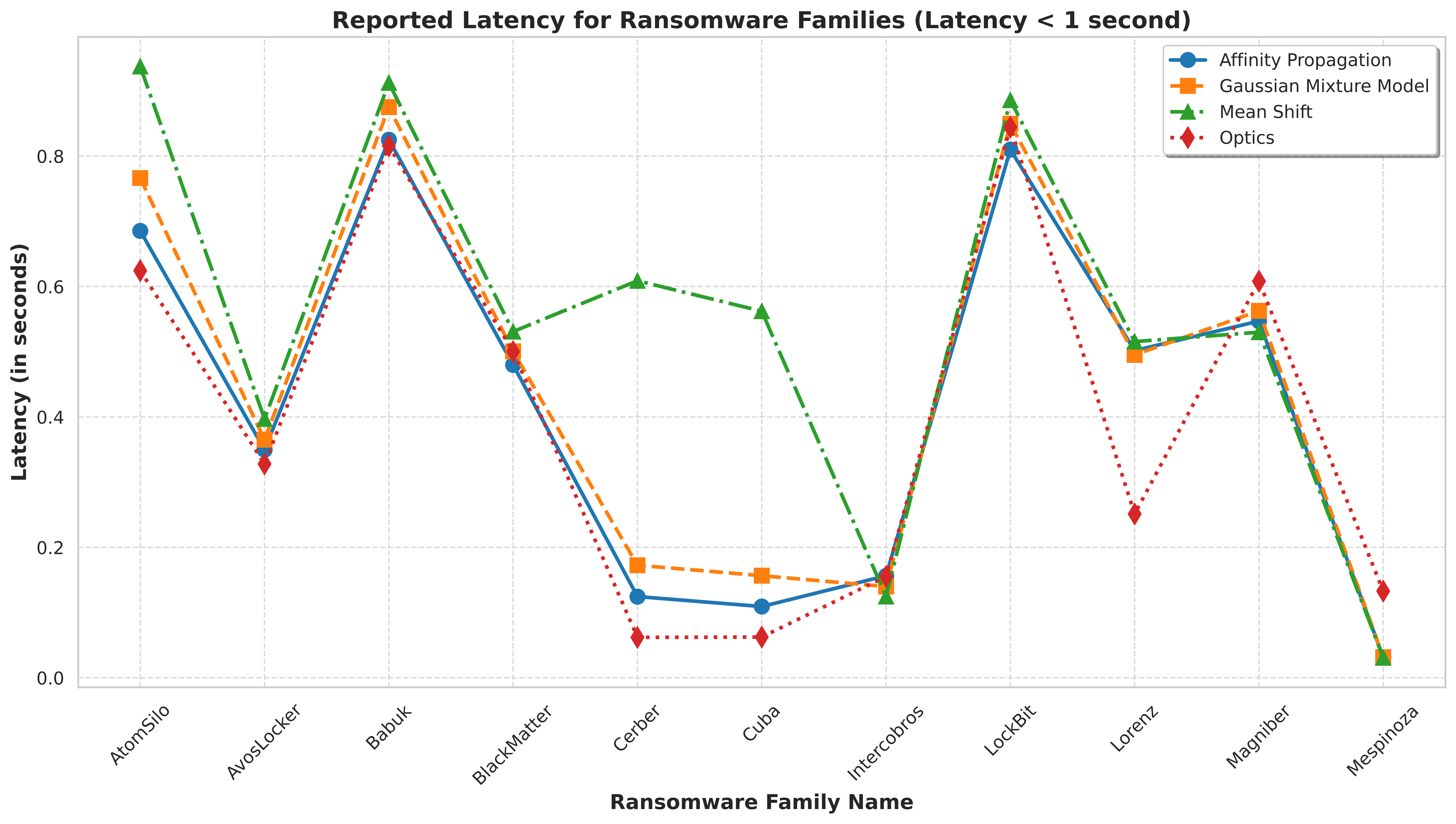}
        \caption{ Reported Latency for Ransomware Families (Latency < 1 second)}
        \label{ddsub1}
    \end{subfigure}
    \vspace{1em} 
    \begin{subfigure}[t]{\textwidth}
        \centering
    \includegraphics[width=0.80\textwidth,height=70mm]{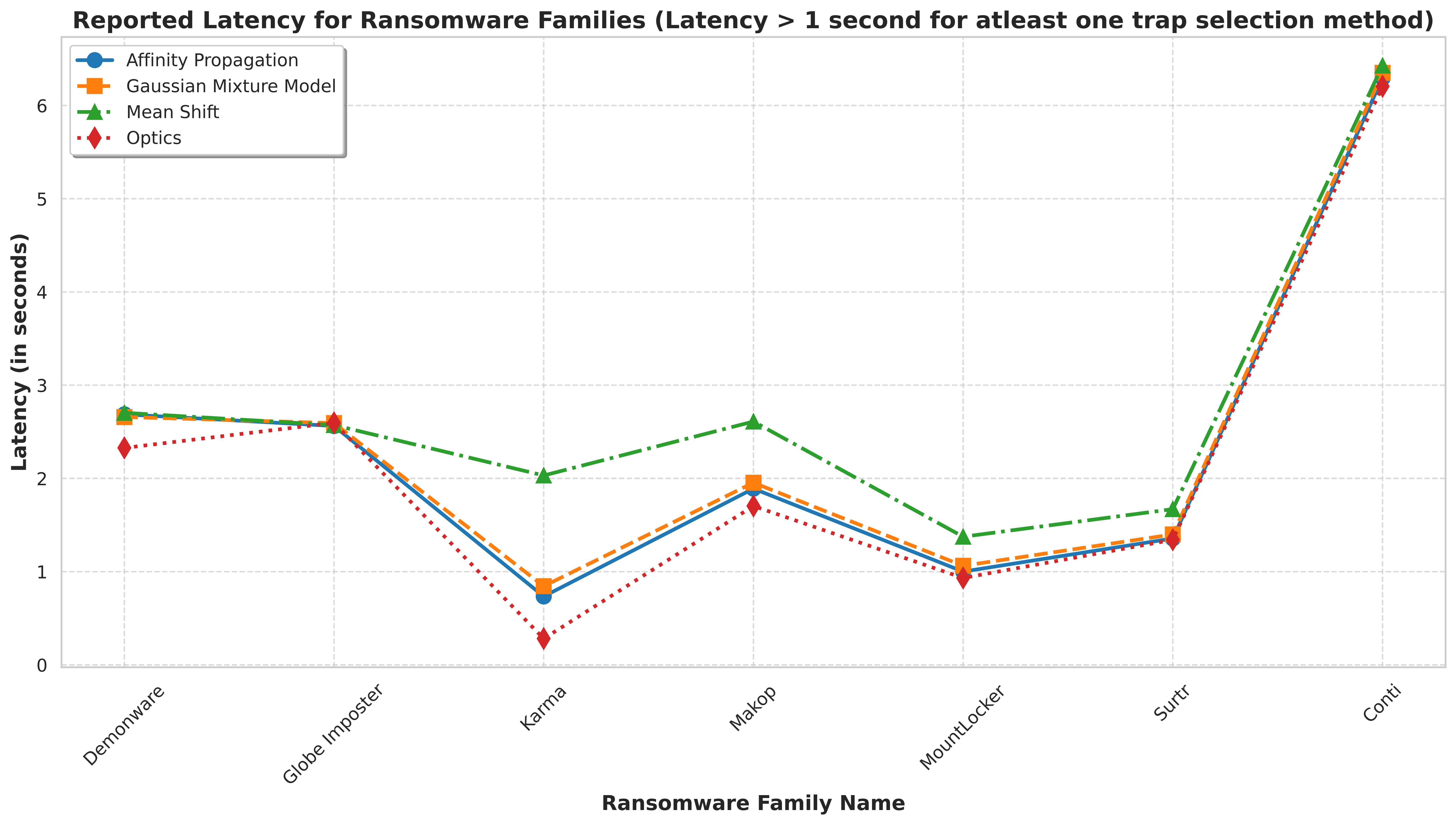}
        \caption{Reported Latency for Ransomware Families (Latency > 1 second for atleast one trap selection method)}
        \label{ddsub2}
    \end{subfigure}
    
    \caption{Comparison of ML based trap selection methods (Criteria: Detection Delay across ransomware variants)}
    \label{DD}
\end{figure} 
The detection delay is defined as the time in seconds from the start of ransomware execution until the monitoring module raises an alert. When ransomware hits any of the selected traps, a timestamp is recorded. The detection delay is calculated as the difference between the time the ransomware execution started and the time the file modification was identified.

Some ransomware variants do not start encryption immediately upon execution \cite{anand2023hiper}. Instead, they first perform activities such as contacting command and control servers, enabling auto-run, stopping anti-malware tools, and scanning the network. These activities introduce a delay before encryption begins. However, with the advent of multithreading, ransomware can encrypt files concurrently with these other activities, allowing it to cause significant damage in a very short time. As a result, some ransomware variants show higher detection delays and lower file loss because they perform multiple actions before starting file encryption. Conversely, other variants exhibit minimal detection delays and cause higher file loss due to their use of multithreading and efficient encryption algorithms.


Figure \ref{DD} illustrates the detection delay observed across multiple ransomware variants. The variants of Conti, DemonWare and GlobeImposter showed higher detection delays (shown in Figure \ref{ddsub2}), while Cerber, Cuba, AvosLocker, and Intercobros showed lower detection delays (shown in Figure \ref{ddsub1}). We also observed that the majority of ransomware families in Category-2 (mentioned in Table-\ref{RFAM}) resulted in a higher detection delay, because they launch fewer threads to carry out encryption. While detection delay is majorly influenced by the efficiency of trap selection methods and the pre-encryption duration, the multi-threaded nature of ransomware also plays a significant role. When ransomware launches parallel threads to encrypt files across multiple directories, the time it takes to trigger the selected traps is reduced, leading to faster encryption and potentially quicker detection. However, with fewer threads, this process is slower, contributing to the increased detection delay. Notably, AvosLocker exhibited a detection delay of less than 0.5 seconds, yet it caused the loss of more than 500 files. This behavior highlights the rapid encryption speed of AvosLocker compared to other variants.  In terms of average detection delay, the results correlate with the extent of file loss. The traps generated by Optics resulted in the lowest delay of 1.10 seconds, followed by Affinity Propagation with a 1.17 second delay, GMM with a 1.2 second delay, and Mean Shift with a 1.4 second delay.


\subsection{Utilization of system resources performance impact}

\begin{figure}[h!]
 \centering
 \includegraphics[width=0.5\textwidth, height=60mm]{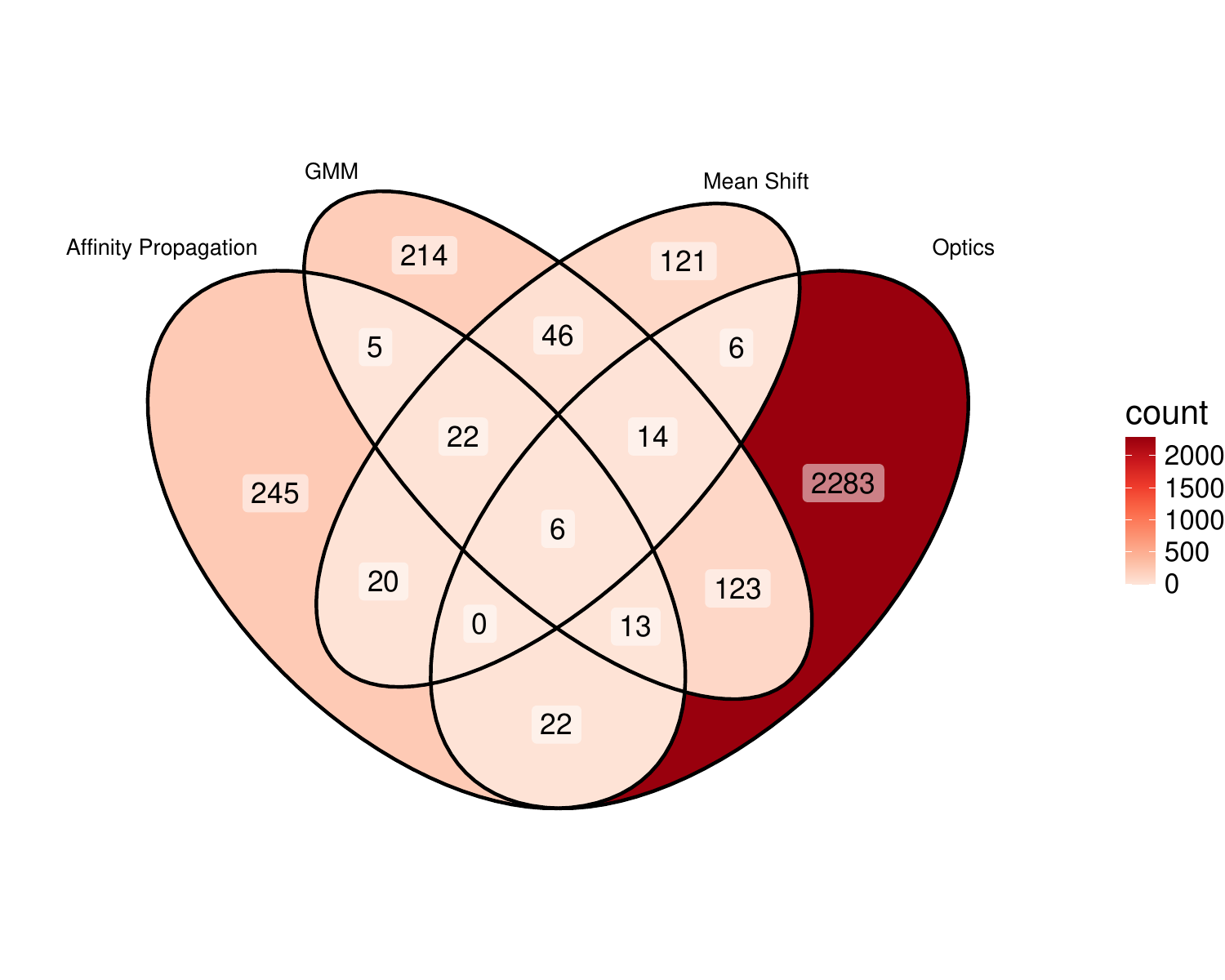}
 \caption{Venn Diagram on the Trap files selected by 4 non-parametric clustering methods}
 \label{4venn}
\end{figure}

We select existing files on the endpoint as traps and continuously monitor them for any potential modifications. As we select more files for monitoring, the memory usage for the monitoring process increases. Out of the 20,558 files present on the endpoint, Affinity Propagation selected 333 files, GMM selected 443 files, Mean Shift selected 235 files, and Optics selected 2283 files as traps. All the trap selection methods differ significantly in their clustering processes, which is evident in the selection of traps. As shown in Figure \ref{4venn}, only six files are common across all the methods. Optics selects 11.10\% of the files as traps, the highest utilization of user files, while Mean Shift selects only 1.15\% of user files, the lowest among the four trap selection methods. Optics selects more files as traps due to its capability to identify data points with varying densities, as detailed in Section 3.4 . Optics works by varying the density threshold, allowing it to find clusters even in areas with relatively low point density, leading to the identification of more clusters. More clusters lead to an increase in cluster centers, which are utilized as traps in our approach. The number of files selected as traps affects runtime memory utilization. The monitoring process consumes 56.29 MB to continuously monitor the files selected by Optics, the highest among all methods, while it consumes only 21.26 MB for the files selected by Mean Shift, the lowest.

\subsection{Comparison of ML based Trap selection methods and its limitations}


\begin{table}[htp]
\centering
\resizebox{\textwidth}{!}{%
\begin{tabular}{ccccc}
\hline
\textbf{Algorithm Name} &
  \textbf{\begin{tabular}[c]{@{}c@{}}Percentage of files \\ selected as Traps\end{tabular}} &
  \textbf{\begin{tabular}[c]{@{}c@{}}Average File Loss \\ Reported (in Number)\end{tabular}} &
  \textbf{\begin{tabular}[c]{@{}c@{}}Average Detection Delay \\ Reported (in Sec)\end{tabular}} &
  \textbf{\begin{tabular}[c]{@{}c@{}}Average Load on \\ the Machine (in MB)\end{tabular}} \\ \hline
Affinity Propagation & \color[HTML]{0000FF}1.633                       & \color[HTML]{0000FF}133.33 
& \color[HTML]{0000FF}1.1765                      & \color[HTML]{0000FF}22.3                       \\
GMM                  & 2.172                       & 155.5 
& 1.2059                      & 24.4                       \\
Mean Shift           & \color[HTML]{38761D} 1.152  & \color[HTML]{FF0000} 239.55 
& \color[HTML]{FE0000} 1.4019 & \color[HTML]{036400} 21.26 \\
Optics               & \color[HTML]{FF0000} 11.105 & \color[HTML]{38761D} 90.16 
& \color[HTML]{38761D} 1.1034 & \color[HTML]{FF0000} 56.29 \\ \hline
\end{tabular}%
}
\caption{Comparison of ML based Trap selection algorithms}
\label{COMPTAB}
\end{table}


 ``A good trap selection method should minimize the number of traps required to detect ransomware, thereby reducing file loss and minimizing performance overhead on the endpoint". Although Optics provided better results in terms of detection delay and file loss, it selected a larger number of files as traps, causing a higher load on the endpoint, as shown in Table- \ref{COMPTAB}. Since one of the objectives of this work is to suggest a method that can be deployed at endpoints to continuously monitor for ransomware activity, Optics is not an ideal choice. Mean Shift selects fewer files as traps, resulting in a lighter load on the endpoint. However, it has the highest file loss and detection delay, making it unsuitable for trap selection. Among all the methods we explored, Affinity Propagation stands out as the second best in terms of file loss and detection delay, with a minimal average load of 22 MB on the endpoint.  This makes \textit{Affinity Propagation better suited for trap selection compared to the other methods}.

\begin{table}[htp]
\centering
\resizebox{0.80\textwidth}{!}{%
\begin{tabular}{|c|c|cccc|}
\hline
\multirow{2}{*}{\textbf{Endpoint ID}} & \multirow{2}{*}{\textbf{File Distribution: Average number of files per folder}} & \multicolumn{4}{c|}{\textbf{File Loss \%}} \\ \cline{3-6} 
            &                   & \multicolumn{1}{c|}{\textbf{AP}} & \multicolumn{1}{c|}{\textbf{GMM}} & \multicolumn{1}{c|}{\textbf{MS}} & \textbf{Optics} \\ \hline
EP-1 & 447 files/folder  & \multicolumn{1}{c|}{0.6523}      & \multicolumn{1}{c|}{0.7602}       & \multicolumn{1}{c|}{1.1749}      & 0.4423          \\ \hline
EP-2 & 1064 files/folder & \multicolumn{1}{c|}{0.9415}      & \multicolumn{1}{c|}{1.0519}       & \multicolumn{1}{c|}{1.3635}      & 0.6917          \\ \hline
\end{tabular}%
}
\caption{File loss comparison of ML based Trap selection algorithms on two end point machines}
\label{ENDPOINTCOMP}
\end{table}

\begin{figure}[htp]
    \centering
    \begin{subfigure}[t]{\textwidth}
        \centering
        \includegraphics[width=0.75\textwidth, height=65mm]{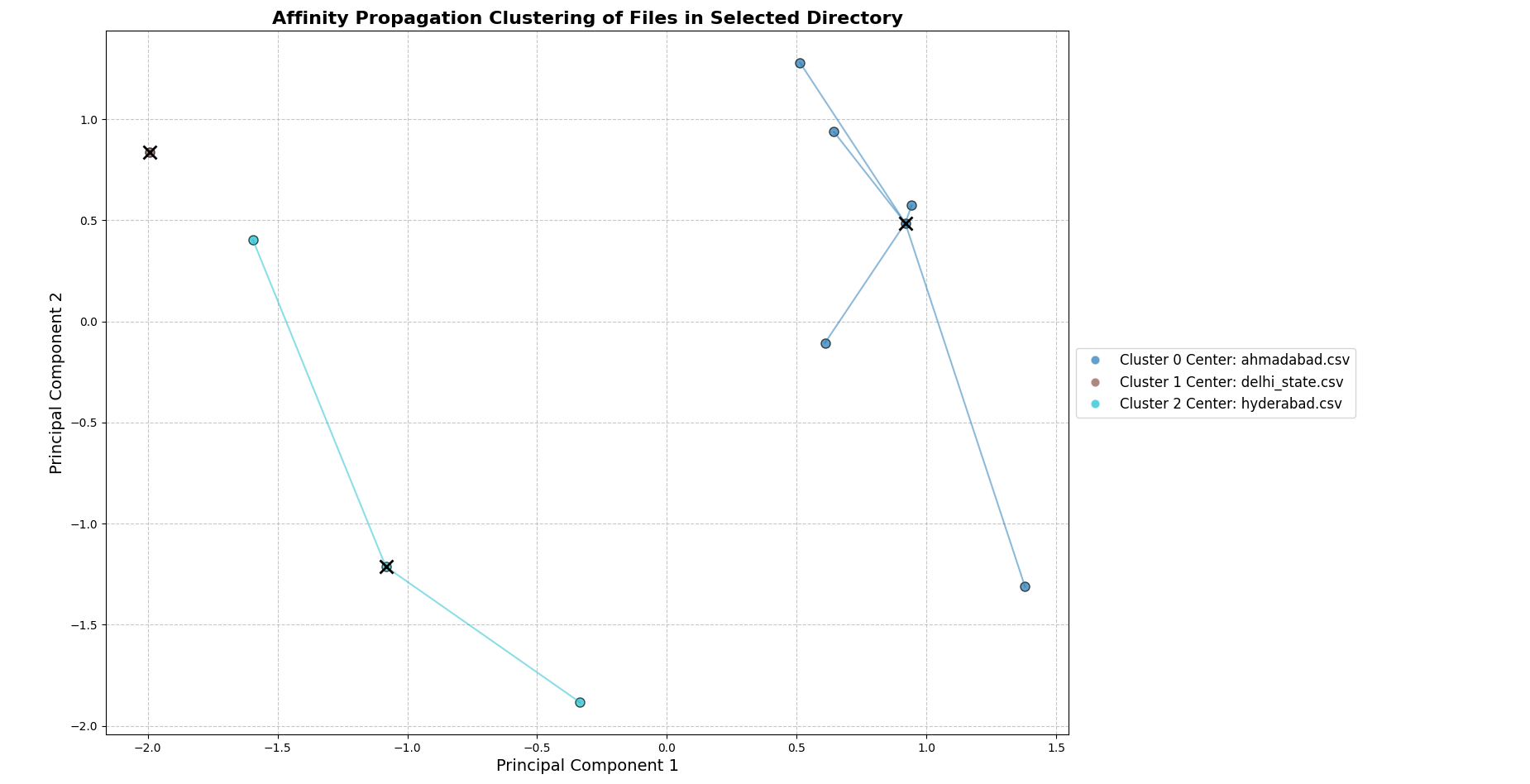}
        \caption{ Cluster formation on a directory with fewer files}
        \label{5sub1}
    \end{subfigure}
    
    \vspace{1em} 
    \hspace{10em}
    \begin{subfigure}[t]{\textwidth}
        \centering
        \includegraphics[width=0.75\textwidth,height=65mm]{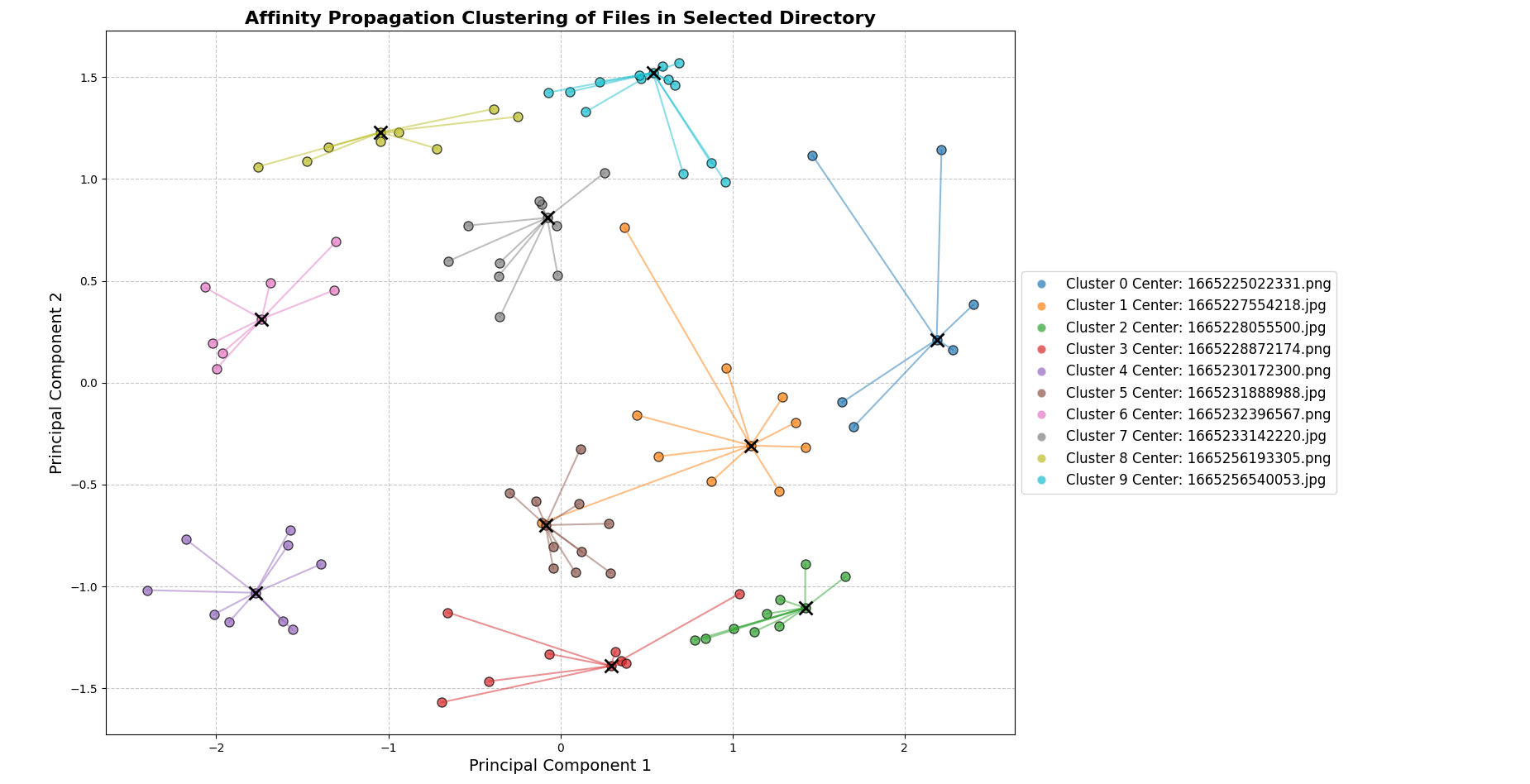}
        \caption{Cluster formation on a directory with more files}
        \label{5sub2}
    \end{subfigure}
    
    \caption{Cluster formation: Two different scenarios}
    \label{ClusterPLOT}
\end{figure}
To verify our results, we conducted an evaluation on another endpoint system with the same configuration. However, we organized the files differently so that the average number of files per directory varied between the two endpoints. Endpoint 1 (EP-1) is the machine described in the section -\ref{sec4ep} with fewer files in each directory, while Endpoint 2 (EP-2) contains more files in each folder. We observed a similar pattern of results across both endpoints, as shown in Table-\ref{ENDPOINTCOMP}. Notably, we found that as the number of files per folder increased, the percentage of file loss also increased.

\textbf{\underline{Limitations}}:
The increase in file loss percentage when the number of files per folder is increased can be attributed to the file ordering schemes used by modern ransomware variants. During trap file selection, we consider parameters such as file size, creation time, modification time, and file type to combat ransomware that encrypts files based on recent access logs, file type, or size or random order. However, the file name order (whether alphabetic or reverse alphabetic) is not taken into account when clusters are created using the ML-based trap selection methods.

From our analysis, we noticed that ransomware groups, such as AvosLocker, Conti, and BlackMatter, encrypt files in reverse alphabetic order. In contrast, Babuk, AtomSilo, and LockBit follow an alphabetic order. The following scenarios elucidate the reasons for file loss when the ransomware adheres to the file name order of encryption:
\begin{enumerate}
    \item \textbf{Scenario 1 (Directory with less number of files)}: Figure \ref{5sub1} illustrates trap file selection in a directory with a fewer number of files using the Affinity Propagation algorithm. Out of 10 files, 3 are identified as traps and are monitored in real-time for ransomware detection. Even if the ransomware variant employs multi-threading to encrypt files and follows any file name ordering, it will be detected when it modifies any one of the three selected traps. In this scenario, file loss is minimal since traps are selected from every directory.
    \item \textbf{Scenario 2 (Directory with more number of files)}: Figure \ref{5sub2} illustrates trap file selection in a directory with more number ($\ge$ 100) of files using the Affinity Propagation algorithm. Out of 100 files, 10 are identified as traps and are monitored in real-time for ransomware detection. Since the traps are selected without considering file name order, they could be located anywhere in the sorted list of the file encryption order. If ransomware encrypts files in alphabetic or reverse alphabetic order, the monitoring module will raise an alert with a considerable delay. This delay results in higher file loss, as shown in Table \ref{ENDPOINTCOMP}.
\end{enumerate}

\subsection{Proposed Method for Trap Selection : APFO}

    
    

\begin{figure}[htp]
    \centering
    \begin{subfigure}[t]{0.48\textwidth}  
        \centering
        \includegraphics[width=\textwidth, height=60mm]{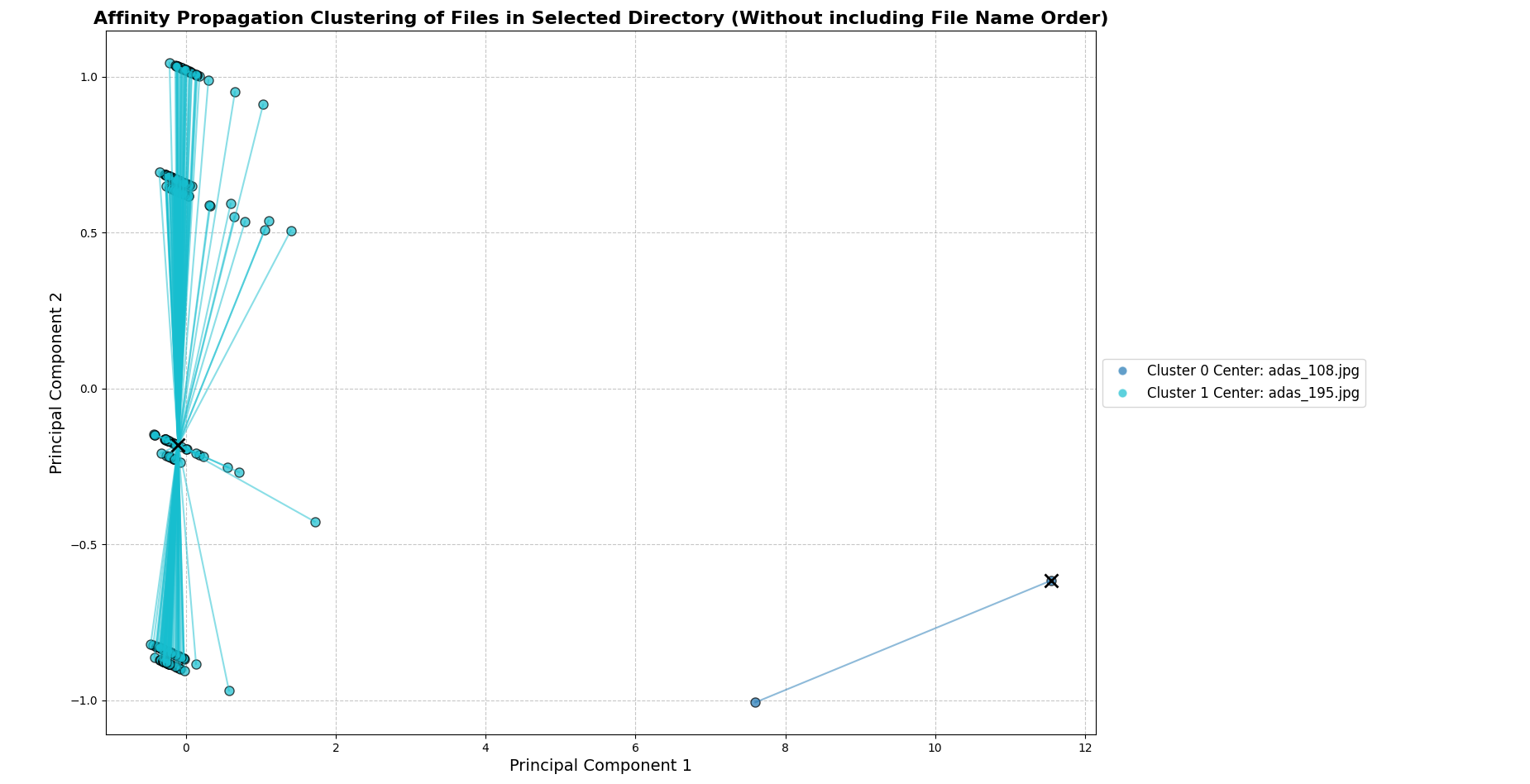}  
        \caption{Cluster formation without including File Name Order Features (Dir 1)}
        \label{7sub1}
    \end{subfigure}
    \hfill 
    \begin{subfigure}[t]{0.48\textwidth}
        \centering
        \includegraphics[width=\textwidth, height=60mm]{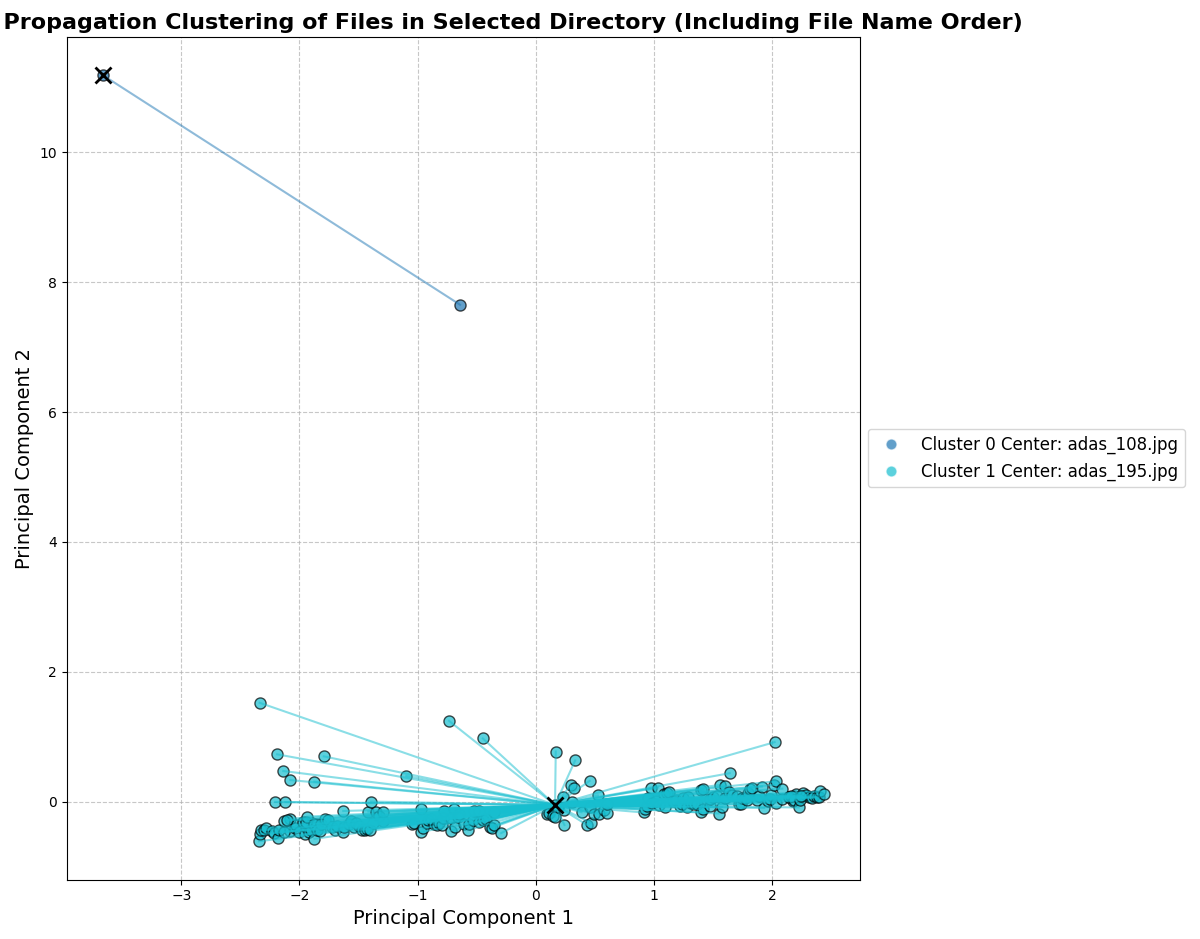}  
        \caption{Cluster formation including File Name Order Features (Dir 1)}
        \label{7sub2}
    \end{subfigure}
    
    \vspace{1em} 
    
    \begin{subfigure}[t]{0.48\textwidth}
        \centering
        \includegraphics[width=\textwidth, height=60mm]{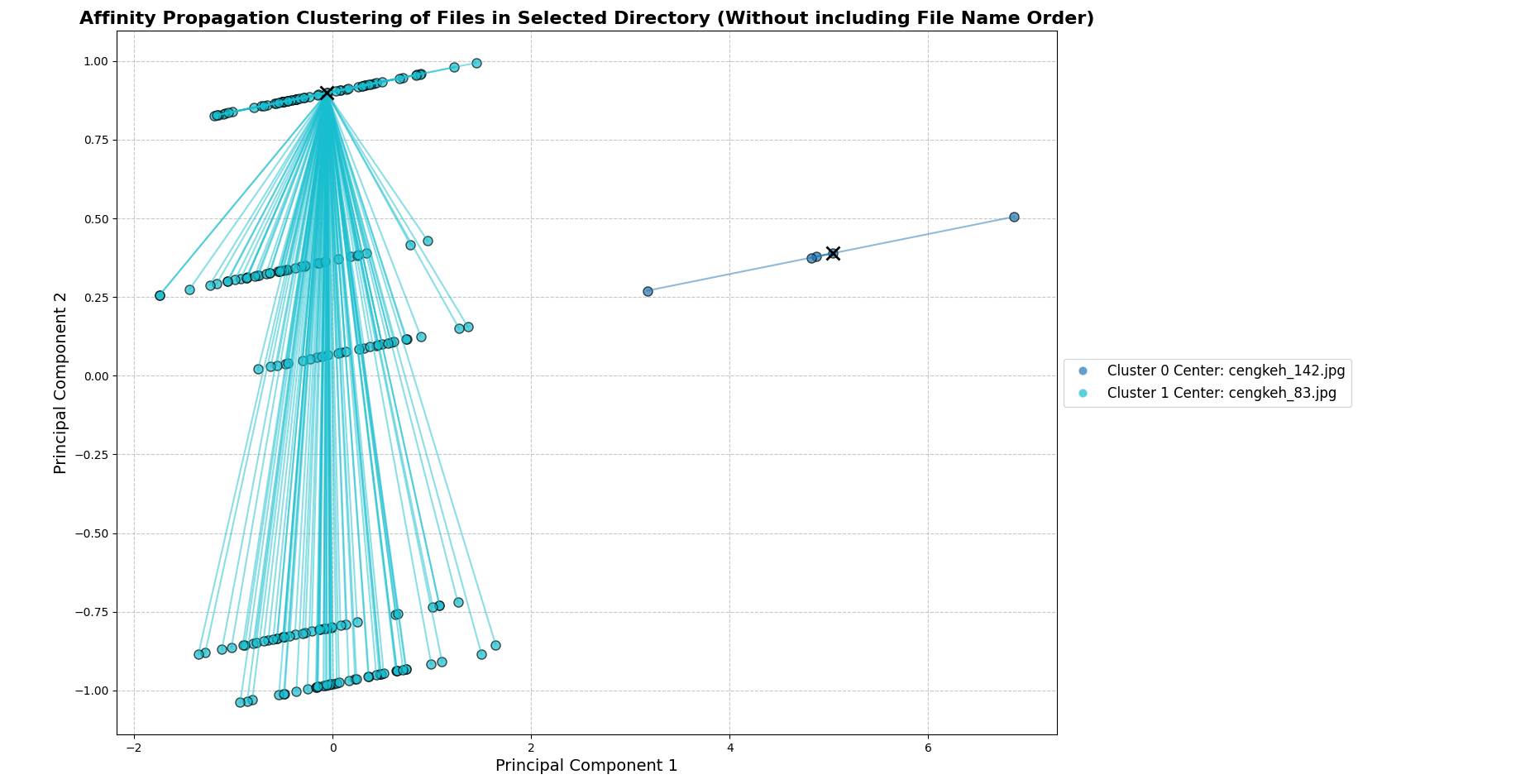}  
        \caption{Cluster formation without including File Name Order Features (Dir 2)}
        \label{7sub3}
    \end{subfigure}
    \hfill
    \begin{subfigure}[t]{0.48\textwidth}
        \centering
        \includegraphics[width=\textwidth, height=60mm]{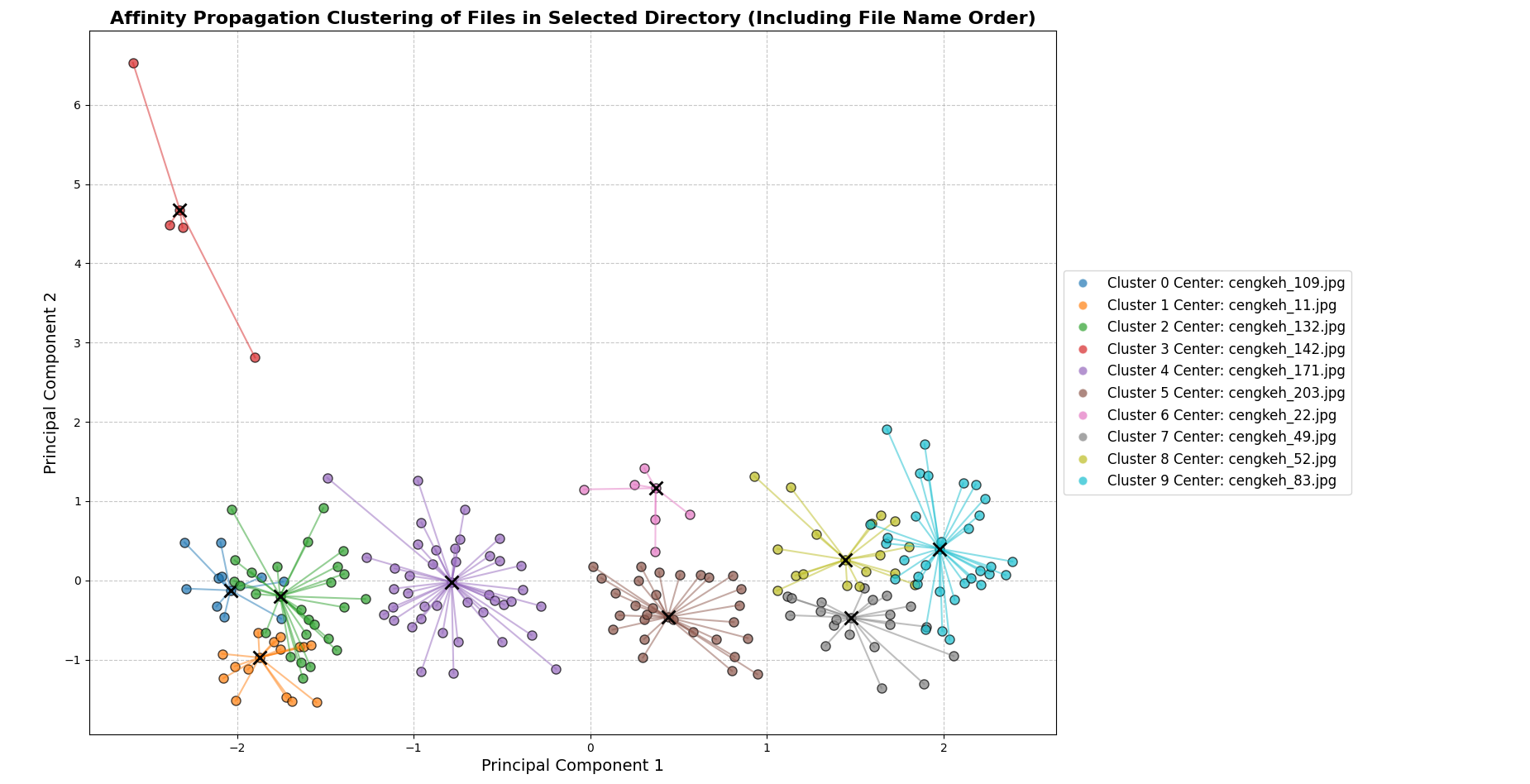}  
        \caption{Cluster formation including File Name Order Features (Dir 2)}
        \label{7sub4}
    \end{subfigure}
    
    \caption{Cluster formation with and without File Name Order Features: Two different directories}
    \label{FOFig}
\end{figure}

After researching different methods, we found that Affinity Propagation is the best machine learning-based trap selection algorithm when it comes to reducing file loss, detection delay, and keeping performance overhead to a minimum value. The authors of the RTrap approach also employed Affinity propagation for trap selection, aligning with our findings \cite{ganfure2023rtrap}. While RTrap works well for early detection of ransomware that encrypts files based on recent access logs, file type, size, or random order, it has some limitations. RTrap, in particular, does not account for scenarios in which traps are selected from directories containing more than 100 files, resulting in delayed detection and increased file loss when ransomware encrypts files in alphabetical or reverse alphabetical order.

To reduce the gap, we integrated file name order features—specifically, alphabetical and reverse alphabetical order—into our existing feature set for trap selection. During the feature extraction phase for a given directory, we sorted the files alphabetically and applied the standard scalar method to normalize the feature vector, ensuring a zero mean and unit variance. In this process, the first file in alphabetical order is assigned a value of 1, while the last file is assigned a value of -1. Similarly, We also extracted feature vectors for files in reverse alphabetical order and combined these new features with the existing feature set for clustering.

For experimentation, we selected two directories (Dir 1 \& Dir 2) containing 210 files each and evaluated clustering under two scenarios:
\begin{enumerate}
    \item Clustering with the directory's feature set, excluding file name order features.
    \item Clustering with the directory's feature set, including file name order features.
\end{enumerate}

For the first directory, as shown in Figures \ref{7sub1} and \ref{7sub2}, both scenarios—clustering with and without file name order features—resulted in the selection of same trap files. When ransomware encrypts this directory in alphabetical order, it could result in the loss of 108 files, whereas encryption in reverse alphabetical order might result in the loss of 15 files. However, for the other directory, shown in Figures \ref{7sub3} and \ref{7sub4}, including file name order features led to a better selection of traps compared to trap selection without these features. Even in this scenario, file loss would be around 11 files if ransomware follows alphabetical order and 8 files if it follows reverse alphabetical order during encryption.

This observed inconsistency suggests that adding file name order to the list of features, along with file size, type, creation, and modification times, does not always provide a significant advantage. Although we included alphabetical and reverse alphabetical order features, these two features alone do not directly affect the outcome of clustering and exemplar selection. Other features, such as file size, type, creation time, and modification time, also play crucial role in trap selection. Therefore, incorporating file name order into the feature set does not consistently offer a substantial benefit when dealing with ransomware that follows the alphabetic or reverse alphabetic order of encryption.

\begin{figure}[htp]
 \centering
 \includegraphics[width=\textwidth,height=85mm]{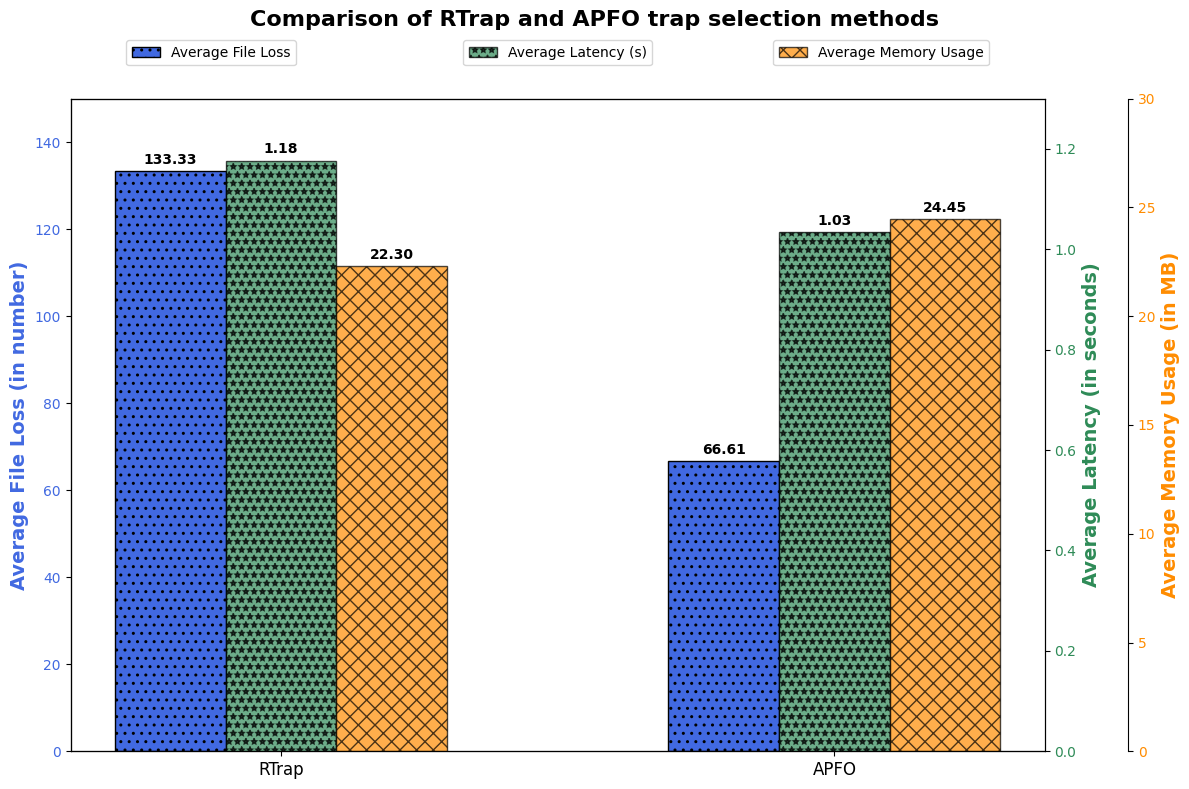}
 \caption{Comparison of RTrap and APFO file trap selection methods}
 \label{APFO}
\end{figure}

To overcome this limitation and reduce file loss, we propose APFO (Affinity Propagation with File Order). APFO combines machine learning-based trap selection with a heuristic approach. The heuristic method selects files in both alphabetical and reverse alphabetical order across all user directories, while Affinity Propagation selects files based on non-parametric clustering, considering parameters such as file size, creation and modification times, and file type. The combined list of selected files, excluding duplicates, is monitored for ransomware detection.

Figure \ref{APFO} illustrates the comparison between APFO and RTrap file selection methods. With APFO, file loss decreases significantly from 133 to 67 files, and detection delay is reduced from 1.18 seconds to 1.03 seconds in comparison with RTrap. This improvement is largely attributed to the ransomware families AvosLocker and Babuk, which encrypt files in reverse alphabetical and alphabetical order, respectively. While APFO does result in slightly higher memory consumption compared to RTrap due to the monitoring of a combined list of trap files, it significantly improves detection delay and reduces file loss compared to all other machine learning-based trap selection methods. APFO can be deployed as an agent on Windows machines to continuously monitor for ransomware without disrupting other system activities.

\section{Conclusion and Future Work}
Using machine learning for trap selection to detect ransomware is an emerging area of research. Most related works focus on heuristic-based methods for selecting trap files, which often fail when ransomware employs multi-threading and random encryption orders. In this study, we explore non-parametric clustering methods, including Affinity Propagation, GMM, Mean Shift, and Optics, to select traps at endpoints. We evaluate these selected traps to identify the most effective method that minimizes file loss and detection delay without causing significant performance overhead. Our analysis reveals that Affinity Propagation is the most suitable machine learning method for trap selection and can be easily deployed at endpoints compared to other methods. It resulted in a minimal file loss of 0.62\% on 20,558 files, with a performance overhead of 22.3MB.

However, ransomware variants like Babuk and AvosLocker, which follow alphabetical and reverse alphabetical file encryption orders, can significantly delay machine learning-based trap selection methods, leading to substantial file loss. To address this issue, we propose APFO—a combination of Affinity Propagation with file name order. APFO leverages both the machine learning-based trap selection method of Affinity Propagation and the heuristics of alphabetical and reverse alphabetical file ordering to select files across directories. APFO performed significantly better than Affinity Propagation alone, resulting in a very minimal file loss of 0.32\%, a detection delay of 1.03 seconds, and a performance overhead of 24.5MB on a given endpoint. This makes APFO highly effective for deployment at endpoints to counteract even faster ransomware encryption methods. Future work will focus on pre-encryption scenarios of ransomware execution, aiming to identify and learn behavioral patterns to further improve early detection.


\bibliographystyle{IEEEtran}
\bibliography{references}

\end{document}